\documentclass[aps,prl,groupedaddress,twocolumn]{revtex4}
\usepackage{graphicx}
\usepackage{amssymb}
\usepackage{wasysym}
\usepackage{pifont}
\usepackage{caption}

\def\frac12{{1\over 2}}
\newcommand{\bra}{\langle}
\newcommand{\ket}{\rangle}
\newcommand{\be}{\begin{equation}}
\newcommand{\ee}{\end{equation}}
\begin{document}

\title{Universal Protein Distributions in a Model of Cell Growth and Division}

\author{Naama Brenner}
\affiliation{Department of Chemical Engineering and Laboratory of Network Biology, Technion, Haifa 32000, Israel}
\author{C.M.~Newman}
\affiliation{Courant Institute of Mathematical Sciences, New York, NY 10012 USA and NYU-ECNU Institute of Mathematical Sciences at
NYU Shanghai, 3663 Zhongshan Road North, Shanghai 200062, China}
\author{Dino Osmanovi\'c}
\author{Yitzhak Rabin}
\affiliation{Department of Physics and Institute of Nanotechnology and Advanced Materials,
Bar-Ilan University, Ramat Gan 52900, Israel}
\author{Hanna Salman}
\affiliation{Department of Physics and Astronomy, Department of Computational and Systems Biology, University of Pittsburgh, Pittsburgh, PA 15260 USA}
\author{D.L.~Stein}
\affiliation{Department of Physics and Courant Institute of Mathematical Sciences,
New York University, New York, NY 10012 USA and NYU-ECNU Institutes of Physics and Mathematical Sciences at NYU Shanghai, 3663 Zhongshan Road North, Shanghai, 200062, China}

\begin{abstract}
Protein distributions measured under a broad set of conditions in bacteria and yeast were shown to exhibit a common skewed shape, with variances depending quadratically on means. For bacteria these properties were reproduced by temporal measurements of protein content, showing accumulation and division across generations. Here we present a stochastic growth-and-division model with feedback which captures these observed properties. The limiting copy number distribution is calculated exactly, and a single parameter is found to determine the distribution shape and the variance-to-mean relation. Estimating this parameter from bacterial temporal data reproduces the measured distribution shape with high accuracy, and leads to predictions for future experiments.
\end{abstract}


\maketitle

\section{Introduction}

The phenotype of a biological cell --- in particular, the types and
copy numbers of its expressed proteins --- fluctuates from cell to
cell, even among those whose genotypes and growth environments are
identical (reviewed in~\cite{Maheshri07,Raj08,Sanchez13}).  Protein
content depends on a complex interplay of genetic, epigenetic and
metabolic processes, with numerous cell-specific regulatory mechanisms
and feedback loops. However,
recent experiments~\cite{Salman12} have demonstrated that for two
different types of microorganism (yeast and bacteria), each under a
broad range of conditions, the distribution of highly expressed
protein copy number appears {\it universal\/}: under rescaling by mean
and standard deviation, all such distributions collapse onto a single
skewed curve~\cite{noteRNA}. In the same experiments variances were found to depend
quadratically on their means,
a trend displayed also
by all highly expressed proteins in~{\it E.~coli\/} in a genome-wide
study~\cite{Taniguchi10} (see
also~\cite{Furasawa2005}).

A recent study following the protein content in individual {\it E.~coli\/} bacteria over roughly 70~generations has
revealed that, under the same scaling criteria, the shape of the
distribution of protein copy number sampled over time in an individual
converges to the one observed in large
populations~\cite{Brenner15}. While analogous temporal data are
currently unavailable for yeast, this is an important property that
reflects the ergodicity of the relative fluctuations in protein
expression in bacteria.  These results can serve as a basis for
constructing a model relating bacterial temporal protein dynamics to
their distributions. 

\section{Model Without Feedback}

Given that the universal statistical properties
described above were found for a range of experimental conditions for
various proteins in bacteria~\cite{Salman12}, such a model should rely
only on general coarse-grained processes. We therefore start by
assuming as little as possible given the experimental data:

\begin{itemize}

\item  Protein number increases as $e^{k_it}$
during the $i^{\rm th}$ generation, where the exponential growth rate $k_i$ fluctuates
with $i$~\cite{Brenner15}.

\item The time $T_i$ of the $i^{th}$ generation (i.e., the time between cell division at
the $(i-1)^{\rm st}$ generation and that at the $i^{\rm th}$) is also random \cite{Powell63,Biswas2014A,Osella2014,SRBA14,Jun15}.

\item The product $X_i=k_iT_i$ is a random variable, with
(positive) mean $\mu$ and variance $\sigma^2$.  We will refer to $X_i$ as the {\it accumulation exponent\/}.

\item Protein number is conserved at cell division, and protein degradation is much slower than a typical interdivision time~\cite{notefast}.

\end{itemize}

Let $N_i$ denote the copy number of a given type of protein in the cell and
$f_i$ the copy number ratio between the daughter and parent cells, both at
the end of the $i^{\rm th}$ generation.  Incorporating the features listed
above gives rise to the recursion relation
\begin{equation}
\label{eq:rate}
N_{i+1}=f_iN_i\exp{(X_{i+1})}\, .
\end{equation}
In bacteria $f_i$ is narrowly distributed about~1/2~\cite{Mannik12}. We take $f_i=1/2$ for
now and discuss deviations from this assumption later. The solution
of~(\ref{eq:rate}) for arbitrary generation number $n$ is
\begin{equation}
\label{eq:soln}
N_n= 2^{-n}N_0\exp\Bigl({\sum_{j=1}^nX_j}\Bigr)
\end{equation}
where $N_0$ is the initial copy number. Then
\begin{equation}
\label{eq:log}
\ln(N_n/N_0)=-n\ln2+\sum_{j=1}^nX_j
\end{equation}
The mean of~$X_j$ should compensate for the decrease in protein number in the daughter cells caused
by division;
otherwise copy numbers would be unstable, running off to unsustainably large numbers or falling to zero within a few generations.
Eq.~(\ref{eq:log}) can be rewritten as
\begin{equation}
\label{eq:final}
\Bigl[\ln(N_n/N_0)-n(\mu-\ln 2)\Bigr]/\sqrt{n}\sigma = \Bigl[\sum_{j=1}^nX_j-n\mu\Bigr]/\sqrt{n}\sigma\, .
\end{equation}
If the accumulation exponents $X_j$ are independent, the central limit
theorem gives
\begin{equation}
\label{eq:CLT}
\Bigl[\ln(N_n/N_0)-n(\mu-\ln{2})\Bigr]/\sqrt{n}\sigma\rightarrow {\cal{N}}(0,1)\, ,
\end{equation}
that is, the LHS converges in distribution to the normal distribution with
mean zero and variance one.  There is no stationary distribution for this process; the mean and variance
of $N_n$ vary with time (or $n$), even when $\mu=\ln 2$
exactly.  This conclusion holds independently of the various distributions used;
all that matters is that fluctuations are independent between generations. This analysis demonstrates that a stationary distribution, as experimentally observed,
can result only if some negative feedback is present. Given this, we next introduce and analyze a modified model with effective feedback regulating protein accumulation, and following that we discuss its experimental justification and consequences.

\section{ Model with Feedback}

A given protein type in an individual cell
has a well-defined typical copy number $\overline{N}$. Its value is
nonuniversal, depending on protein type, growth conditions and possibly
other biological factors~\cite{Brenner15,noteN}. The stationary
distribution shape must therefore be independent of $\overline{N}$.

A natural extension of the growth-and-division model consistent with
observations is the introduction of
an accumulation exponent
that is
negatively correlated with protein number at the start of the cycle.
The experimental requirement of universality constrains the form of
the feedback term: a change in scale of $\overline{N}$ cannot alter
the functional form of the recursion relation. The only function with
this property of scale invariance is the power law; the modified
recursion relation is therefore
\begin{equation}
\label{eq:modified}
N_{i+1} = f_iN_i[\exp(\xi_{i+1})](N_i/\overline{N})^{-\alpha}
\end{equation}
with $f_i$ defined as before, $\xi_{i+1}$ (which we will call the residual accumulation exponent) the component of the
accumulation that fluctuates independently from generation to
generation, 
and a new phenomenological parameter $0 < \alpha < 1$;
$\alpha=0$ is the case without feedback~\cite{notealpha}.

The recursion relation of the modified model is
\begin{equation}
\label{eq:univ1}
\ln N_{i+1} = \ln f_i +\xi_{i+1} + (1-\alpha)\ln N_i+\alpha\ln\overline{N}.
\end{equation}
It is not hard to check that first, there is now a limiting stationary distribution, with $\bra N\ket\approx\overline{N}$; and second, that $\overline{N}$ can be scaled out of the growth equations.

The introduction of a nonzero $\alpha$ makes the specific form of the limiting distribution dependent (though not too sensitively; see below)
on the distributions of $f_i$ and $\xi_i$.
Experiments on bacteria indicate that $\xi_i$ is approximately normally distributed (see Fig.~2b).
Using this, we can solve for the limiting distribution exactly when the division ratio is fixed. The limiting distribution is again lognormal:
\begin{eqnarray}
\label{eq:lognormal}
P(N) = {1\over N\Sigma\sqrt{2\pi}}
\times\exp\Bigl[-{\Big(\ln N-{\cal{M}}\Big)^2\over 2\Sigma^2}\Bigr],
\end{eqnarray}
\noindent with ${\cal{M}}=\ln\overline{N}+(\mu -\ln 2)/\alpha$ and
$\Sigma=\sigma/\sqrt{2\alpha-\alpha^2}$. These two parameters together
determine the mean and variance of the distribution: specifically, $\bra N
\ket=\exp\{{\cal{M}}+\Sigma^2/2\}$, and $\bra N^2 \ket-\bra N
\ket^2=(e^{\Sigma^2}-1)\exp\{2{\cal{M}}+\Sigma^2\}$. However, only $\Sigma$
determines the shape of the distribution; different values of ${\cal{M}}$
collapse on one another following scaling by a linear
transformation.
Moreover, for fixed $\Sigma$ the variance scales
quadratically with the mean.

In terms of the model, Eq.~(\ref{eq:lognormal}) has several important
features. First, it preserves universality under scaling with respect to
all variables that appear only in the parameter ${\cal{M}}$, because the
distribution shape is independent of this additive term. Consequently, all
values of $\overline{N}$, $\mu$, and division ratio yield the same
distribution shape.

Second, as noted above a single composite parameter $\Sigma$ determines the shape of the distribution. $\Sigma$ characterizes the balance between the {\it variance} of accumulation exponents, which tends to drive the process to diverge, and the effective feedback parameter $\alpha$, which provides a ``restoring force''. Once $\Sigma$ is determined, both properties -- collapse of scaled distributions and quadratic dependence of variance on mean -- are preserved.

Third, it should be noted that in the setting of our model, the limiting steady state distribution equally well represents the time average over many generations of a single individual or the average at a single large time over a large population in which the individuals are evolving independently.

The analysis above assumed a Gaussian distribution of the accumulation exponents and a fixed value of the division ratio. We now explore the robustness of our conclusions.
Without these assumptions, the lognormal solution will no longer be exact, but will not be significantly altered for a variety of unimodal distributions for both variables. Moreover, the scaling properties within classes characterized by $\Sigma$ (defined as before and which is now close to but not exactly equal to the standard deviation of $\ln N$) still hold.
Fig.~1a shows examples of means and variances computed from many
simulations, in which the $f_i$'s were drawn from a Gaussian
distribution and the $\xi_i$'s from a gamma~distribution. For each
simulation, $\alpha$ and $\overline N$ were chosen randomly, and the
variance $\sigma^2$ of the gamma~distribution was adjusted to give a
shape parameter $\Sigma$ equal to one of three values: 0.2, 0.4 or
0.6. The resulting means and variances are seen in Fig.~1a to
collapse onto three parabolas corresponding to the three classes
defined by the value of $\Sigma$. Limiting distributions are shown in
Figs.~1(b-d);
while they are all very close to lognormal, the different $\Sigma$'s lead to a range of shapes, from nearly Gaussian (1b; $\Sigma=0.2$), to skewed with exponential-like tail (1c; $\Sigma=0.4$) and finally to highly skewed~(1d; $\Sigma=0.6$). Within each class the distributions from all simulations collapse after rescaling onto a single curve. Thus, both properties of distribution collapse and the quadratic dependence of variance on mean hold once $\Sigma$ is fixed, even though the conditions of the exactly solvable model are relaxed.

In addition, the assumption of symmetric division can be relaxed as long as the average accumulation compensates for the loss at division.
The assumption of stable proteins can also be relaxed to include first-order protein degradation;
this would require an additional parameter and would modify only the accumulation exponent.

\begin{figure}[h]

               \includegraphics[width=80mm]{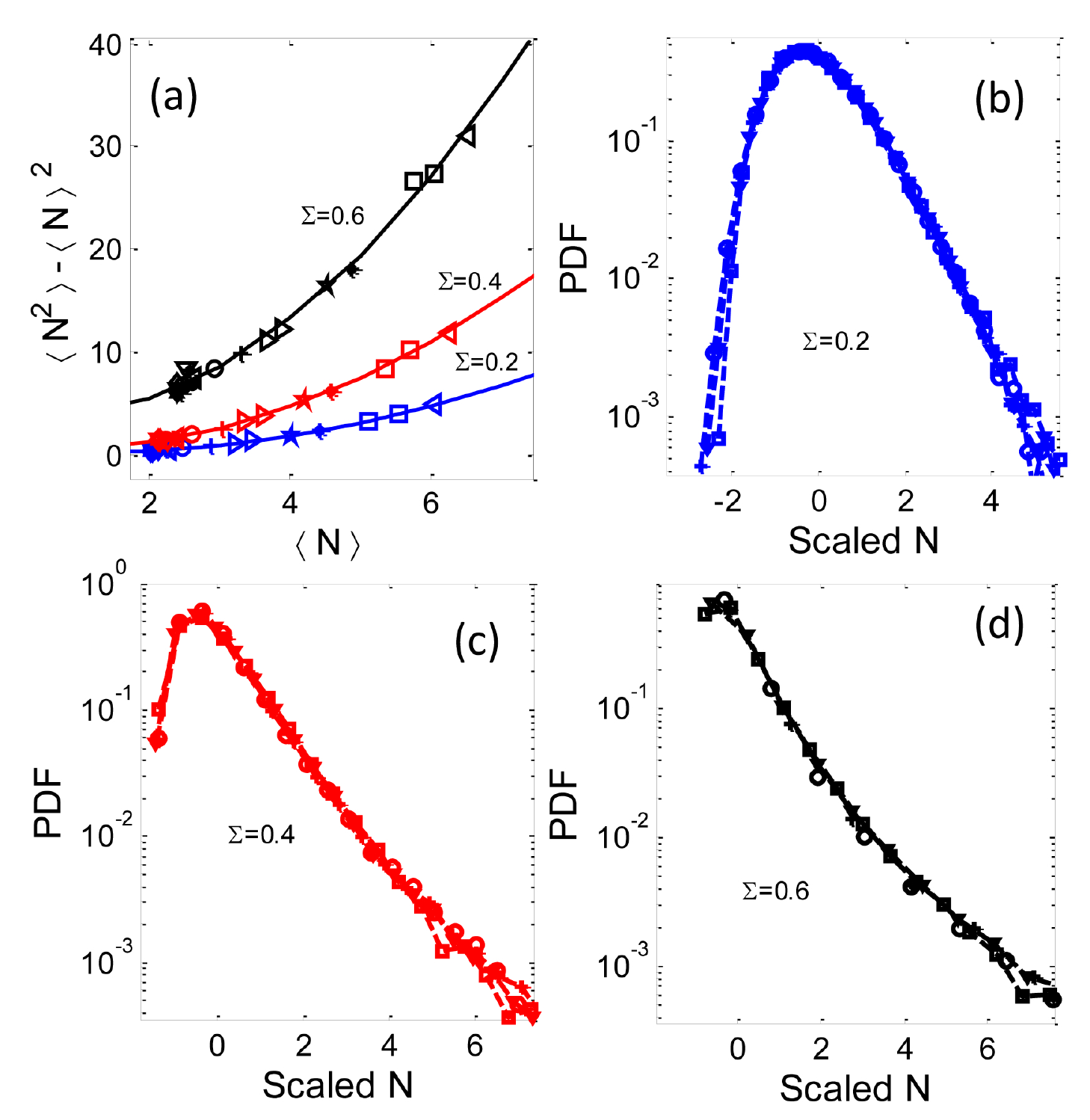}
               \vskip -4mm

                \caption{{\bf Model simulations.} The stochastic process described by Eq.~(\ref{eq:modified}) was simulated over 20,000 generations for each run. Values for $\alpha$ and $\overline N$ were chosen uniformly at random in the range [0,1] and [0.25,0.75], respectively. The division ratio $f_i$ was drawn from a Gaussian~distribution with mean 1/2 and standard deviation 0.1, and $\xi_i$ from a gamma~distribution whose variance was adjusted to $\alpha$ to obtain one of 3 values of the shape parameter $\Sigma$.  (a) Variance vs.~mean of 18 simulations for each shape parameter show a collapse on 3 corresponding parabolas. (b-d) Limiting distributions of three simulations for each class are all well fit by lognormal but have different shape parameters and so span a range of shapes, from approximately Gaussian (b), to an exponential-like tail (c), to a highly skewed distribution (d). In each class the distributions collapse onto one another to high accuracy. }
    \label{fig:Model}
    \vskip -2mm
\end{figure}

\section{Comparison with Data}

To test the assumption of negative correlation we plotted experimental values of $\Delta\ln N_i=\ln N_{i+1}-\ln N_i$ vs.~$\ln N_i$, as measured for bacteria, in Fig.~2a.
The data points were collected from six individual trajectories normalized to unit average (data from~\cite{Brenner15}).
\begin{figure}[h]
 \includegraphics[width=80mm]{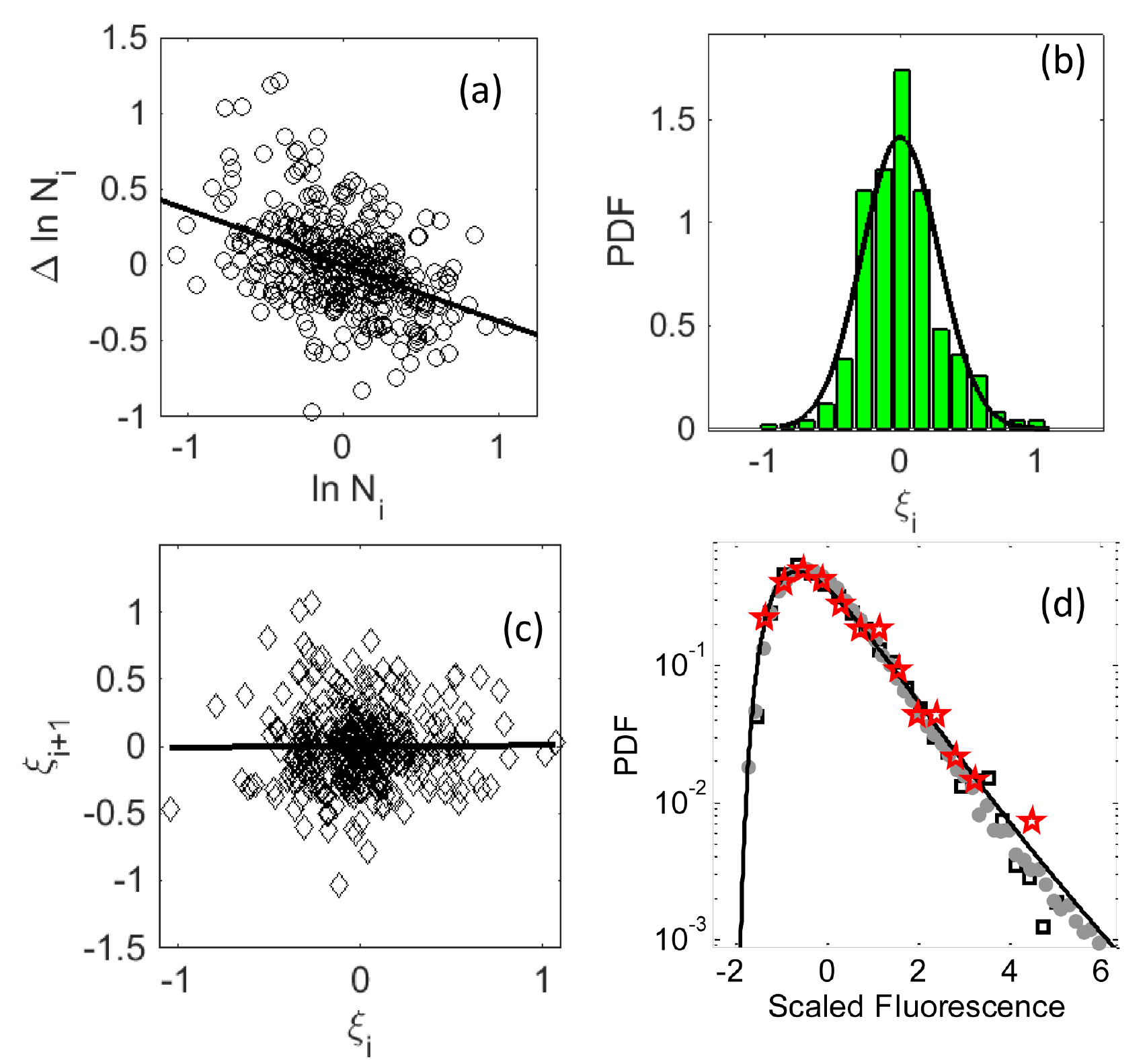}
               \vskip -4mm
                \caption{{\bf Comparison with data.} (a)  $\Delta\ln N_i\equiv\ln N_{i+1} -\ln N_i$ plotted vs.~$\ln N_i$ in units of $\overline{N}$. Solid line is $y=-0.37x$. (b) Accumulation exponents in consecutive generations are  approximately normally distributed with average~ln2 (subtracted out in the figure), and (c) are independent between generations. Solid line is $y=0.0075x-0.00036$. (d) Estimating the universality class parameter~$\Sigma$ for these data from (a) and (b), the distribution shape is predicted by Eq.~(\ref{eq:lognormal}) (black line) and compared to data. Grey circles: large population snapshot; black squares: protein trajectories of individual trapped bacteria;  red stars: sampled points at the end of each cell cycle.  Data from~\cite{Salman12,Brenner15}.  }
    \label{fig:AlphaEst}
    \vskip -2mm
\end{figure}
In agreement with Eq.~(\ref{eq:univ1}), the data are consistent with random scatter about an overall linear dependence, with negative slope determining $\alpha$ to be approximately~$0.37\pm 0.04$. Using this value the
residual
accumulation exponents~$\xi_i$ can be extracted from the data using Eq.~(\ref{eq:univ1}) and measured values of $N_i$. The approximately Gaussian distribution of these exponents is shown in Fig.~2b, and their independence between consecutive generations is evident from Fig.~2c.

The parameter determining the distribution shape in our model is
$\Sigma=\sigma/\sqrt{2\alpha-\alpha^2}$. Estimating $\alpha$ and $\sigma$
from Figs.~2a and~2b respectively, we find $\Sigma\approx
0.4\pm 0.02$.  Fig.~2d shows the lognormal distribution of Eq.~(\ref{eq:lognormal}) corresponding to this parameter in rescaled units (black line), together with data from a large bacterial population~(grey circles), and single cell trajectories (black squares and red stars) that exhibits the measured universal distribution shape over several decades of probability. We note that this is not a fit, but a model prediction with no adjustable parameters: the single parameter determining the distribution shape is computed separately using the single-cell dynamic measurements.

\section{Discussion}

Motivated by experiments that found universal protein
distributions under various conditions in yeast and bacteria, and by
single-cell measurements of protein accumulation and division in bacteria
across multiple generations, we have presented a model based on the premise
that the combined processes of growth, division, and feedback set the
distribution shape. With fixed division ratio and Gaussian randomness the model is exactly
solvable.  The solution identifies a single parameter
$\Sigma$~(Eq.~(\ref{eq:lognormal})) defining the distribution shape: it
quantifies the balance between growth of variance and feedback that
stabilizes protein numbers. With $\Sigma$ fixed, a rescaling by mean and
standard deviation collapses these distributions onto a single curve, and
displays a quadratic relation between variance and mean.

Thus, our model predicts that populations in the same class
--- i.e., which share the same shape parameter ---
exhibit similar
set-point balance between the opposing forces in the dynamics of their
protein content across time,  i.e. between the {\it variance} of accumulation exponents ($\sigma$) that drive the process to diverge, and the feedback parameter ($\alpha$) that prevents divergence. Therefore, if the variance of the
exponents $\xi_i$ changes, the feedback parameter $\alpha$ should change in a
correlated manner. To test this prediction, single-cell dynamical
trajectories need to be measured over a variety of conditions that span
these parameters. Another possibility consistent with our model is that
both $\sigma$ and $\alpha$ are fixed. At the moment, experimental
perturbations --- for example, changing medium or temperature --- can change
the mean, for example by modifying the mean cell cycle time; but their effect on the variance of exponents is
unknown~\cite{Biswas2014A}.

Our approach shares some features with previous theoretical work but
differs in other respects.  Earlier work focused on protein accumulation
and division~\cite{Berg1978,Brenner2006,Brenner2007,Friedlander2008,Yomo2011}, or
protein accumulation and continuous
dissipation~\cite{Paulsson2005,Friedman2006}.
The recent data on
protein content over multiple generations~\cite{Brenner15} shows that, due to the
exponential nature of protein accumulation, division or dissipation alone
cannot stabilize copy numbers, and reveal a correlation between variables
across generations.

The classes of proteins of interest are those consisting of high-copy-number molecules, 
characterized by exponential accumulation between 
successive cell divisions.
The exponential accumulation of protein during a cell cycle suggests that protein
production reflects a coherent integration of many correlated
processes in the cell. Exponential growth of the
cell size between divisions, as well as negative correlation
analogous to the one reported here, were measured in several recent
experiments~\cite{Osella2014,Campos2014,Biswas2014A,Jun15}. Moreover,
results on trapped bacteria show explicitly that the exponents of cell
size growth and protein accumulation are strongly correlated on a
cycle-by-cycle basis~\cite{Brenner15}. This suggests a picture where
highly expressed proteins that are strongly coupled to cell metabolism
are components of multi-dimensional phenotypes that covary between
individual cells.  This view is supported by a model recently proposed
to explain exponential biomass growth as resulting from an interacting
network of reactions~\cite{Biswas2014B,Pugatch15}. Furthermore, our model is mathematically related to a recently proposed model of cell size
regulation~\cite{SRBA14,Amir14}, which finds under similar assumptions
a lognormal distribution with the same compound parameter governing
its shape. For highly expressed proteins, this may be expected since protein production and cell growth are tightly coupled~\cite{Brenner15}.
However, there are also important differences between the two models, which we address in detail in the Appendix. 

Our model addresses directly the universal behavior of bacterial protein distribution among different biological realizations,  including expression regulation mechanisms, growth conditions, and types of microorganism.
Its ingredients are independent of specific biological mechanisms
and rely 
on those general aspects of cellular events ---
exponential protein accumulation,
division and feedback --- 
that are likely to be common to all dividing cell populations. 
This marks a significant departure from the main current
line of research on protein number variation, which investigates
synthetically produced proteins while experimentally isolating the
contribution of specific microscopic
mechanisms~\cite{BarEven06,Murphy2007,Hornung2012,Carey2013,Dadiani2013,Sharon2014}.

In particular, we have observed that
feedback must be present, because without it the mean and variance
necessarily drift to larger values as time increases.  Moreover,
regardless of the specific processes leading to feedback (which may differ
for different protein types and organisms), the mathematical form of
the feedback in a growth-and-division model must be power law to be
consistent with universality.
 \vskip 4mm
\begin{acknowledgments}
{\bf Aacknowledgments} We thank Erez~Braun for many useful discussions.  This work was supported by the US-Israel Binational Science Foundation (NB and HS).
DLS thanks the John Simon Guggenheim Foundation for a fellowship that
partially supported this research, and Bar-Ilan University for its hospitality while this work was initiated.
YR acknowledges support by the I-CORE Program of the Planning and Budgeting committee and the Israel Science Foundation, and by the US-Israel Binational
Science Foundation. NB acknowledges support of the Israel Science Foundation (Grant No. 1566/11). HS acknowledges the support of NSF Grant PHY-1401576.
CMN acknowledges the support of NSF Grant DMS-1007524 and DLS acknowledges the support of NSF Grant DMS-1207678.
\end{acknowledgments}

\section{Appendix: Universal Protein Distributions and Cell Growth}

In this appendix, we detail the differences between our model and a mathematically similar model of cell size
regulation recently proposed by Amir ~\cite{SRBA14,Amir14}. However, before describing those differences, we first discuss some nontrivial consequences following from the {\it similarity\/} between the models. The fact that similar
mathematical descriptions, albeit with different biological interpretations, may capture the dynamics of two distinct
phenotypes is a potentially significant mathematical unification resulting from a
strong coupling among disparate biological processes. As noted above, cell size and protein copy number 
are two separate, but strongly correlated, phenotypes of the cell. 
Which one controls the other, or whether both are regulated together, is not known at this time. 
One interesting implication of the work presented here is that not only cell size, but
also the total content of highly expressed proteins, is under the
control of what appears to be a global cellular feedback, supporting
the viewpoint that protein copy number variation is a {\it global\/}
variable. This marks a significant departure
from the main current line of research on protein number variation.

We now turn to a discussion of some important differences in the
interpretation and consequences of the two models.
First, the requirement that the average must scale out of the distribution shape in any model of universal distributions necessitates mathematically the power-law form of the feedback. Second, the parameter $\Sigma$ that governs the ``universality class'' of the family of distributions has been identified, and its constancy leads both to the collapse of all curves under linear scaling and to the observed quadratic relation between variance and mean. The latter is an especially important consequence and has no analogue in cell size distributions. We have also seen numerically that the division into such classes extends beyond the conditions of the analytically solvable case, rendering this result robust with respect to a wide class of distributions of the underlying random variables.

Perhaps most importantly, our model makes specific predictions on the constrained changes allowed in protein trace parameters under varying conditions. Similar predictions cannot currently be made for cell size distributions.

Both our model and that of~\cite{SRBA14,Amir14} are also easily modified to handle asymmetric
division, as in yeast. However, until data are available that relate
temporal to population statistics, it remains to be seen to what
extent the dynamics of proteins across generations in yeast can be
described by the approach outlined in the main text.



\begin{figure}[h]
\includegraphics[width=80mm]{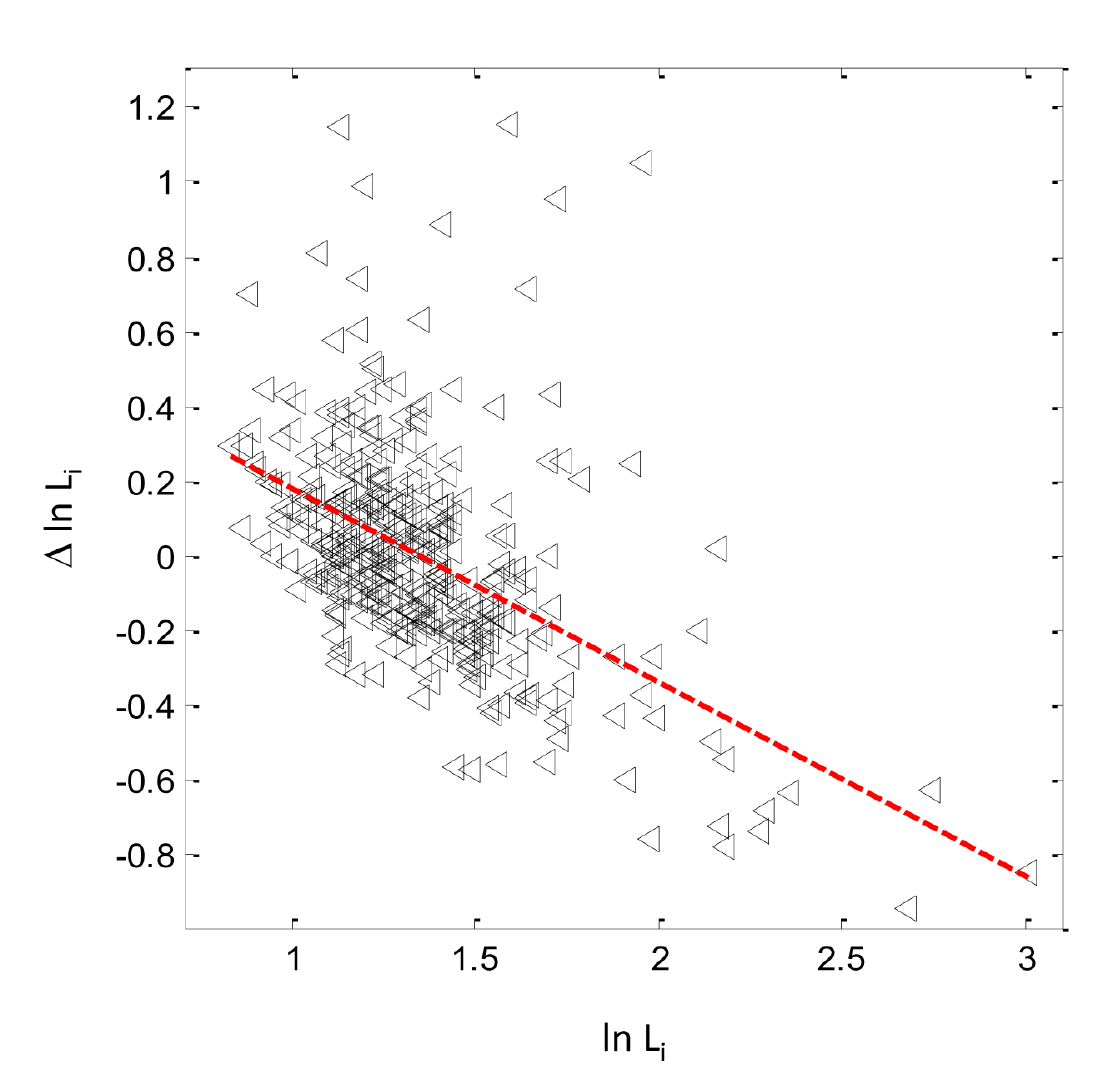}
 \vskip -4mm
 \caption{Cell-size analysis. Similar analysis of that carried out for protein copy number (Fig.~2a) was carried out for cell-size
data,
quantified here by the length $L$ of the bacterial cell in the trapping channel.
$\Delta\ln L_i =\ln L_{i+1} - \ln L_i$, computed from the data in~\cite{Brenner15}, is plotted vs.~$\ln L_i$. The red dashed line is the linear best fit given by
$\Delta\ln L_i  = -0.5185\ln L_i + 0.702$. The slope, which represents the feedback parameter $\alpha$, is approximately 1/2 as predicted by the constant volume addition model described in~\cite{Amir14}.}
    \label{fig:cell}
    \vskip -2mm
\end{figure}

A final important difference between the approach described in this paper and that in~\cite{Amir14} concerns the nature of the feedback
itself. Analysis of E.~coli data led to the conclusion that
cell size feedback is characterized by $\alpha=1/2$~\cite{Amir14},
corresponding (in leading order) to the proposal that the feedback
arises from constant addition of volume over the cell cycle. In
contrast, a different mechanism(s) may apply for copy numbers, and
$\alpha$ can in principle vary among protein types: our current
results are consistent with various values of $\alpha$ with an average
of $\alpha\approx0.37$. However, at this stage there is no direct evidence that can determine which phenotype (cell size, protein copy number, etc.) controls the division point of the cell and thus the feedback mechanism that controls it. In principle, it could also be a cellular state that is defined by several phenotypes simultaneously.

We show this difference explicitly in the figure below. Fig.~\ref{fig:cell} uses cell size 
data from~\cite{Brenner15} to compute the feedback parameter $\alpha$, in manner similar to that used in Fig.~2a,  for the cell size phenotype.
This results in  $\alpha\approx 0.5$, in agreement with~\cite{Amir14}, but different from the value $\alpha\approx 0.37$ shown in Fig.~2a.


\begin{references}
\bibitem{notefast}
A degradation rate that is faster than a typical cell interdivisional time would 
likely simply modify the accumulation exponent, so such proteins would also display exponential increase
of copy number during a cell cycle and be covered equally well by our model. However, we exclude 
such proteins for now to avoid complicating factors.
\bibitem{noteN}
$\overline{N}$ could in principle be modulated on long
timescales by processes not included in our model. If they occur,
such modulations would be sufficiently slow to be decoupled from the fast
growth-and-division processes described here. It is therefore reasonable, and consistent with observations, to
take $\overline{N}$ to be constant over a large number of generations.
\end{references}
\end{document}